\documentclass[twocolumn,showpacs,prl,amsmath,amssymb,epsfig]{revtex4}

\usepackage{graphicx}
\usepackage{dcolumn}
\usepackage{bm}

\begin{document}
\title{Zero-bias anomaly and Kondo-assisted quasi-ballistic 2D
transport}
\author{A. Ghosh, M. H. Wright, C. Siegert, M. Pepper, I.
Farrer, C. J. B. Ford, D. A. Ritchie} \affiliation{Cavendish
Laboratory, University of Cambridge, Madingley Road, Cambridge CB3
0HE, United Kingdom}
\date{\today}

\begin{abstract}
Nonequilibrium transport measurements in mesoscopic quasi-ballistic
2D electron systems show an enhancement in the differential
conductance around the Fermi energy. At very low temperatures, such a
zero-bias anomaly splits, leading to a suppression of linear
transport at low energies. We also observed a scaling of the
nonequilibrium characteristics at low energies which resembles
electron scattering by two-state systems, addressed in the framework
of two-channel Kondo model. Detailed sample-to-sample reproducibility
indicates an intrinsic phenomenon in unconfined 2D systems in the low
electron-density regime.
\end{abstract}

 \maketitle

Zero-bias anomaly (ZBA) in the non-equilibrium characteristics of
ballistic systems provides additional insight into the coupling
mechanism of a conduction electron with its surroundings. For
example, a maximum in the differential conductance ($dI/dV_{\rm SD}$)
at the Fermi energy ($E_{\rm F}$) in case of tunnelling via magnetic
impurities~\cite{kondo_tun}, or through artificially confined quantum
dots~\cite{Kondo_dot_expt,meir1}, has been explained by a Kondo-like
antiferromagnetic coupling of the electron to localized impurity
spin. Similar enhancement observed in clean quantum point contacts in
semiconductor heterostructures has led to much controversy regarding
the spin-structure and effects of lateral confinement in such
systems~\cite{1D_Kondo_expt}. On the other hand, in quasi-ballistic
metallic nanobridges the ZBA shows in a local cusp-like minimum in
$dI/dV_{\rm SD}$ at $E_{\rm F}$, which has been interpreted in terms
of nonmagnetic two-channel Kondo (2CK) framework arising from the
interaction of the electrons with local two-state atomic
defects~\cite{2CK_TLS_expt}. The interest in this lies in the
prediction that in particle-hole symmetric case, such a model flows
to a $T = 0$ fixed point, giving rise to quantum critical behavior
and non-Fermi liquid effects~\cite{2CK_TLS_theory,2CK_matveev}.

Investigation of nonequilibrium ballistic transport in spatially
extended 2D systems is primarily impeded by greater scattering from
background disorder. The small level spacing requires relatively
larger sample dimensions, resulting in significant momentum
relaxation during the course of transport. Even though recent
nonequilibrium studies in mesoscopic 2D electron systems (2DES) have
shown an unexpected ZBA at low background disorder~\cite{self1_ZBA},
the mechanism of transport in this regime remains unclear. Here we
report the experimental observation of a new and unexpected feature
in the nonequilibrium quasi-ballistic 2D transport. At zero magnetic
field in unconfined samples of mesoscopic dimensions, we find a
strong zero-bias enhancement in $dI/dV_{\rm SD}$,  which splits at
very low $T$ ($\lesssim 150$ mK) by a gate voltage ($V_{\rm
g}$)-dependent magnitude. This modifies the linear transport
properties by introducing intermittent low-energy nonmonotonicity in
the linear conductance $G$ ($= dI/dV_{\rm SD}$ at $V_{\rm SD} = 0$,
where $V_{\rm SD}$ is the source-drain bias), as a function of both
$T$ and in-plane magnetic field ($B_{\rm ||}$). At low $V_{\rm SD}$,
an intriguing scaling behavior of $dI/dV_{\rm SD}$ in $T$ is also
observed, which indicates a 2CK-type scattering of electrons in these
systems at low $T$.

The devices were fabricated from Si $\delta$-doped GaAs/AlGaAs
heterostructures. In order to minimize the disorder arising from
Coulomb potential of ionized dopants, we used a thick layer ($\approx
80$ nm) of undoped AlGaAs spacer, and adopted a slow cooling
procedure~\cite{self1_ZBA} to maximize the correlations in the donor
layer. The as-grown mobility of the wafers range over $\mu \sim 1 -
3\times10^6$ cm$^2$/V~s, depending on the donor ($n_{\rm \delta}$)
and electron ($n_{\rm s}$) density. Over the experimental temperature
range, the momentum relaxation rate was typically $\tau^{-1} \lesssim
(0.1 - 1)\times k_{\rm B}T/\hbar$, indicating a quasi-ballistic
nature of the transport~\cite{quasi_ballisticity}. The elastic
scattering length $\lambda \sim v_{\rm F}\tau \sim 10$ $\mu$m, where
$v_{\rm F}$ is the Fermi velocity, provided an upper cutoff to the
device dimensions, restricting the number of elastic scattering
events to very few or none. Measurement of $dI/dV_{\rm SD}$ was
carried out using standard mixed low-frequency ac/dc technique with
ac excitation $\ll k_{\rm B}T$ at each $T$ inside a top-loading
dilution refrigerator with base electron temperature $\approx 32$ mK.

\begin{figure}[h]
\centering
\includegraphics[height=7.349cm,width=7.5cm]{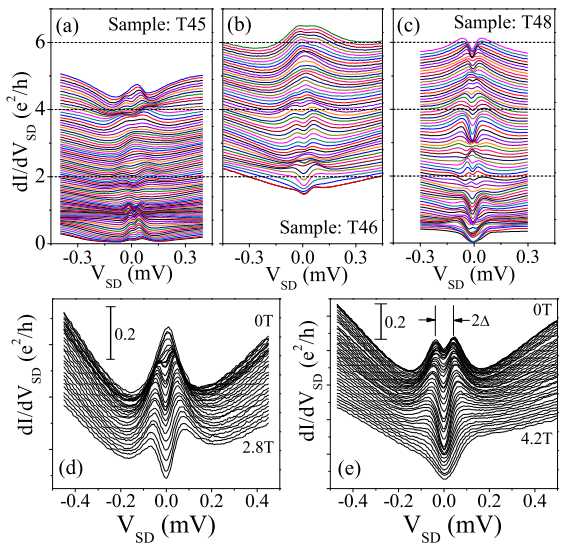}
\caption{(Color online) The zero-bias anomaly at $T \approx 32$ mK
and $B_{\rm ||} = 0$: (a)-(c) Nonequilibrium characteristics in three
different samples. Each differential conductance trace is obtained
for a fixed $V_{\rm g}$. (See text for sample details) (d),(e)
$B_{\rm ||}$-dependence of the nonequilibrium characteristic at the
single peak and double peak regime ($T = 32$ mK). Traces are
vertically shifted for clarity. The vertical bars denote a
conductance change of $0.2\times e^2/h$.}
\end{figure}

Fig.~1(a), (b) and (c) show the $B_{||} = 0$ nonequilibrium
characteristics of three samples with different dimensions and doping
profile. We varied both the total geometrical area, as well as the
aspect ratio in samples T45 ($2\times8 \mu$m$^2$, $n_\delta \approx
2.5\times10^{12}$ cm$^{-2}$), T46 ($5\times8 \mu$m$^2$, $n_\delta
\approx 1.9\times10^{12}$ cm$^{-2}$) and T48 ($5\times5 \mu$m$^2$,
$n_\delta \approx 0.9\times10^{12}$ cm$^{-2}$). The dimensions are
chosen in order to ensure  small single-particle level spacing with
$\Delta\epsilon_{\rm x}$,$ \Delta\epsilon_{\rm y} \sim h^2/8m^*L^2
\ll k_{\rm B}T$ for each $T$. While no consistent feature in $G$ as a
function of $V_{\rm g}$ could be identified (not shown), the ZBA
around $V_{\rm SD} = 0$ is evident for all samples~\cite{disorder}.
We focus on some of the common features of the ZBA shown in Fig.~1:
(1) At $|V_{\rm SD}| \lesssim 0.15-0.2$ meV, $dI/dV_{\rm SD}$ shows
an enhancement at all $V_{\rm g}$. This energy scale varies only
weakly with $V_{\rm g}$, and also $\ll E_{\rm F}$ at all $n_{\rm s}$.
(2) Over certain ranges of $V_{\rm g}$, the ZBA splits by $2\Delta$,
leading to a double-peak structure around $E_{\rm F}$. (3) $\Delta$
is oscillatory in $V_{\rm g}$, and when normalized for series
conductance, in particular for $G \gtrsim 2e^2/h$, the splitting is
strongest when $G$ lies close to an even integral multiple of
$e^2/h$, becoming unresolvable around odd multiples. (4) While no
clear dependence of the energy scales on size/shape of the samples
was observed, the amplitude of the ZBA was found to depend on $n_{\rm
s}$. Note that the ZBA is strongest in T48 ($n_{\rm s} =
6.5\times10^{10}$ cm$^{-2}$ at $V_{\rm g} = 0$), and weakest in T45
($n_{\rm s} = 10.1\times10^{10}$ cm$^{-2}$ at $V_{\rm g} = 0$). Such
weakening is also observed in a given sample as $V_{\rm g}$ (or
$n_{\rm s}$) is increased.

The single- and double-peak regions show distinct behavior in the
presence of external in-plane magnetic field ($B_{||}$). As shown in
Fig.~1(d) and (e), while the single-peak splits by the Zeeman energy
$\Delta_{\rm Z} = 2g^*\mu_{\rm B}B_{||}$ from $B_{||} = 0$, Zeeman
splitting of the double peak appears only at relatively large
(sample-dependent) $B_{||}$-scale ($\sim 0.5 - 2$
Tesla)\cite{self1_ZBA}. Such a splitting of ZBA is taken as a
distinctive feature of Kondo-type
dynamics~\cite{kondo_tun,Kondo_dot_expt}. The effective $g$-factor
$g^*$ was found to be both sample- and $n_{\rm s}$-dependent
$g^*/g_{\rm b} \sim 1 - 3$, where $|g_{\rm b}| = 0.44$, consistent
with exchange-induced enhancement at low $n_{\rm
s}$~\cite{self1_ZBA,g_factor}.

\begin{figure}[t]
\centering
\includegraphics[height=11.15cm,width=7.5cm]{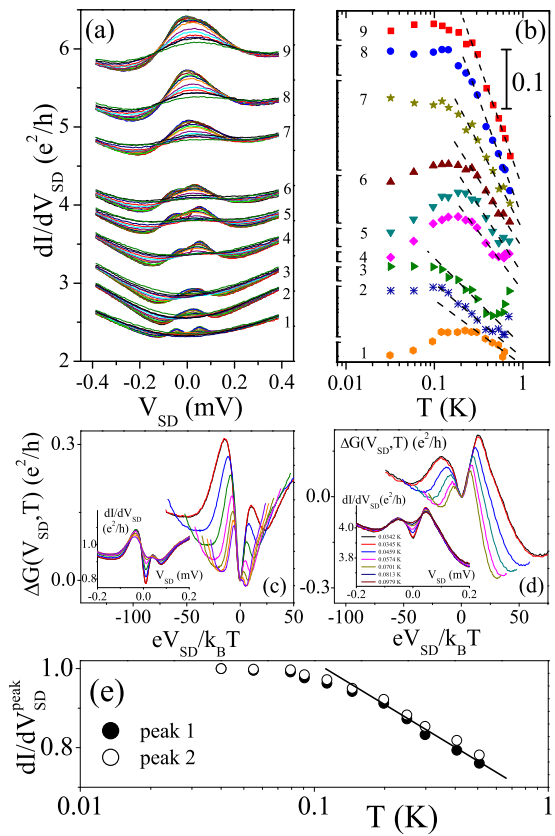}
\caption{(Color online) (a) Suppression of the zero-bias anomaly with
increasing temperature ($T$). Each of the nine sets was recorded at a
fixed gate voltage ($V_{\rm g}$) and $B_{\rm ||} = 0$. (b) Dependence
of the linear conductance $G$ on $T$ at different $V_{\rm g}$. The
length of the vertical bar denote $0.1\times e^2/h$. (c),(d) Collapse
of $\Delta G$ onto a single trace at low-$V_{\rm SD}$ as a function
of $eV_{\rm SD}/k_{\rm B}T$, at two regimes of zero-bias conductance
($G$). The insets show the unscaled original data. (e) $T$-dependence
of individual peaks in a double-peak pattern.}
\end{figure}

Kondo-type dynamics also lead to a suppression of the ZBA with
increasing $T$. Indeed, such a suppression is observed in all our
samples as shown in Fig.~2a for nine different $V_{\rm g}$'s in T46.
In case of single peaks (sets 2, 3, 7 and 8), the suppression is
monotonic as $T$ is increased from $\approx$ 30 mK to $\approx$ 800
mK. For $\Delta \neq 0$, the $T$ dependence of $dI/dV_{\rm SD}$ is
nonmonotonic in $T$ close to $E_{\rm F}$ (see sets 1, 4, 5, 6, and
weakly in 9). Fig.~2b shows the $T$-dependence of $G$ from the traces
in Fig.~2a. Qualitatively, the high-$T$ regime (typically $T \gtrsim
300$ mK) is similar for all cases, where $G$ shows a ``metal''-like
decrease with increasing $T$. For $T \lesssim 100 - 150$ mK, the
increasing behavior of $G$ with $T$ appears intermittently, at values
of $V_{\rm g}$ where $dI/dV_{\rm SD}$ shows resolvable double-peak
structure (Fig.~2b and 3b-e). This reentrant nature is crucial,
because even if the high-$T$ suppression of $G$ is attributed to
various combinations of phonon contribution, the interaction
correction, or the Altshuler-Aronov-type correction from electron
scattering by Friedel oscillations~\cite{int_corr}, the repeated
change in the sign of $dG/dT$ is clearly inconsistent with standard
weak-localization/interaction-based mechanisms having a single
transition from localized to ``metallic'' state transport. However,
in the Zeeman split regime at high $B_{||}$, $G$ increases with $T$
at all $V_{\rm g}$, indicating standard localized state transport.

Most striking aspect of the double-peak structure is the collapse of
$\Delta G(V_{\rm SD},V_{\rm g},T) = dI/dV_{\rm SD}(V_{\rm SD},V_{\rm
g},T) - G(V_{\rm g},T)$, onto a single $V_{\rm g}$-dependent trace
when $V_{\rm SD}$ is scaled by $T$. In Fig.~2(c) and (d) this is
illustrated for ZBA's at two different $G$ (and hence $V_{\rm g}$),
indicating a common underlying mechanism. The insets contain the
actual data prior to substraction of $G(V_{\rm g},T)$, showing the
thermal broadening of $dI/dV_{\rm SD}$, and the suppression of the
ZBA splitting with increasing $T$. In Fig.~2e we show the
$T$-dependence of individual peaks in $dI/dV_{\rm SD}$ in the
double-peak region at the given $V_{\rm g}$. For comparison, the
individual $T$-dependences are normalized by the $dI/dV_{\rm SD}$ at
lowest $T$. Note the approximately logarithmic $T$-dependence of both
peaks at $T \gtrsim 150$ mK, which is also expected in the
Kondo-framework.

The low-$V_{\rm SD}$ scaling of $dI/dV_{\rm SD}$ indicates
nonequilibrium conductance of the form,

\begin{equation}
\label{scale} \frac{dI}{dV_{\rm SD}}(V_{\rm SD},V_{\rm g},T) =
G(V_{\rm g},T) + AT^\alpha {\cal F}\left[V_{\rm g},\frac{eV_{\rm
SD}}{k_{\rm B}T}\right]
\end{equation}

\noindent where $A$ is a phenomenological constant, and $\alpha = 0$
gives the best collapse onto the single sample-dependent function
${\cal F}(V_{\rm g},eV_{\rm SD}/k_{\rm B}T)$. The dependence of
${\cal F}$ on $V_{\rm g}$ will be discussed later. The scaling form
of Eq.~\ref{scale} has been observed in the context of the scattering
of electrons from bistable systems~\cite{2CK_TLS_expt,ralph_PRL}, and
addressed in a two-channel Kondo (2CK)
framework~\cite{2CK_TLS_theory,2CK_matveev}. In quasi-ballistic
nanostructures, 2CK dynamics may arise from two possibilities: (1)
Electron scattering off systems with quasi-degenerate orbital states
acting as pseudospins, while the real spins act as the channel index.
The nature of this orbital degeneracy can however vary from
equivalent lattice defects in metallic nanobridges~\cite{ralph_PRL},
to singlet-triplet degeneracy in multilevel quantum
dots~\cite{kouwenhoven_dot_2CK,Hofs_Scho}. (2) An underscreened
high-spin ($S$) ground state coupled to $M$ ($< 2S$) conduction
channels~\cite{Pust_Glaz}. When parametrized in terms of a
two-impurity Kondo problem, two spins $S_{\rm 1}$ and $S_{\rm 2}$ ($S
= S_{\rm 1} + S_{\rm 2}$) interact with conduction electrons with
antiferromagnetic exchange parameters $J_{\rm 1}$ and $J_{\rm 2}$
($J_{\rm 2} \neq J_{\rm 1}$), and a direct exchange $I$ ($J_{\rm 1},
J_{\rm 2} \gg I$)~\cite{vojta}. A two-stage screening process can
then decouple the spins from the conduction band by forming
$S_1$-$S_2$ singlet if $I$ exceeds some critical magnitude. Since
both stages can be suppressed by lifting the spin degeneracy with
Zeeman energy, a distinctive feature of this mechanism is the
nonmonotonicity of $G$ in both $T$ and $B_{||}$, with a quantitative
correspondence between the respective energy scales~\cite{Pust_Glaz}.

In Fig.~3, we compare the $T$-dependence of $G$ to its
$B_{||}$-dependence in T45 (on a separate cooldown). Four
representative $V_{\rm g}$, with corresponding nonequilibrium traces
at $T \approx 32$ mK and $B_{||} = 0$ T, are identified as $V_{\rm
g1}$ to $V_{\rm g4}$ in Fig.~3a. Figs.~3b-e show the ($B_{||} = 0$)
$T$-dependence of $G$ at these $V_{\rm g}$s, while Figs.~3f-i show
the $B_{||}$-dependence at the base $T \approx 32$ mK. We note that
apart from the qualitative nonmonotonic behavior of $G$ as a function
of both $T$ and $B_{||}$ at $V_{\rm g1}$ and $V_{\rm g3}$, there is
also a quantitative agreement in the energy-scales over which the
double-peak structures are suppressed. For example, as the half width
$\Delta/2$ reduces from $\approx 0.03$ meV at $V_{\rm g1}$ to
$\approx 0.017$ mV at $V_{\rm g2}$, we find a corresponding decrease
in the characteristic thermal and Zeeman energy scales (denoted by
the vertical arrows) from $\sim 0.023$ meV and $\sim 0.028$ meV
respectively at $V_{\rm g1}$, to $\sim 0.011$ meV and $\sim 0.02$ meV
respectively at $V_{\rm g2}$. Since $B_{\rm ||}$ is applied in the
plane of the 2DES (to an accuracy better than $0.2^\circ$),
conventional weak-localization effects are unlikely to contribute to
the observed behavior~\cite{GB_WL}. Coupling of $B_{||}$ to orbital
degree of freedom through spin-orbit interaction ~\cite{GB_WL_SO} or
finite thickness effect~\cite{GB_WL_thick} are also excluded since
neither antilocalization to weak-localization crossover, nor a
stronger suppression of $G$ with increasing $B_{||}$ at lower $n_{\rm
s}$ were observed.

\begin{figure}[t]
\centering
\includegraphics[height=7.085cm,width=7.5cm]{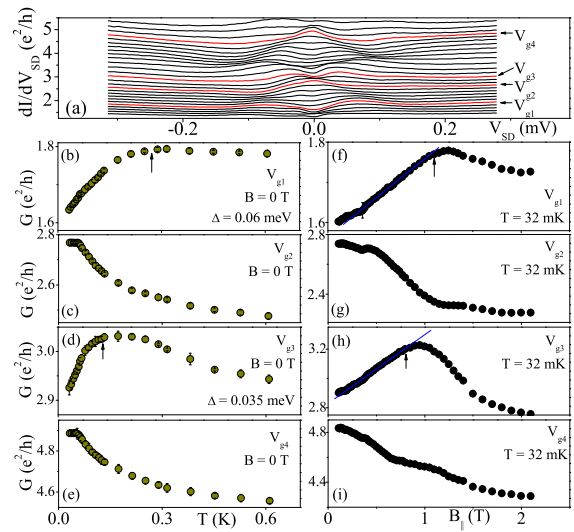}
\caption{(Color online) Correspondence between the source-drain bias,
thermal and Zeeman energy scales: (a) Nonequilibrium traces in T45
identifying two single and two double-peak regions at gate voltages
$V_{\rm g1}$ to $V_{\rm g4}$. (b)-(e) Temperature ($T$)-dependence of
the linear conductance $G$. (f)-(i) In-plane magnetic field ($B_{\rm
||}$) dependence of $G$ at corresponding $V_{\rm g}$s.}
\end{figure}

The Kondo-type enhancement in mesoscopic 2D nonlinear transport in
the ultra-clean limit indicates an unexpected scattering mechanism of
the lead electrons. Moreover, intermittent splitting of the ZBA
implies a two-state nature of the scatterer that becomes resolvable
only at very low $T$ ($\lesssim 150$ mK). In view of the
quasi-ballistic nature of transport, the non-equilibrium conductance
can represent the tunnelling characteristics between the source and
drain across the potential barrier formed by the gate. Presence of
localized magnetic states within the barrier region of traditional
tunnel junctions has been shown to result in Kondo-type enhancement
in the tunnelling conductance~\cite{kondo_tun}. In nanostructures
fabricated from MBE-grown high-quality GaAs/AlGaAs heterostructure,
localized acceptor sites close to the system can significantly modify
the transport through capacitive coupling~\cite{cobden}. This could
lead to a Kondo-assisted tunnelling, similar to that observed in
metallic nanobridges~\cite{buhrman}, and also in semiconductor-metal
junctions~\cite{wolf}. Arguments against such a scenario are, (1)
reproducibility, and insensitivity of the relevant energy scales to
sample-specific details, and (2) enhancement in the amplitude of ZBA
at lower disorder.

\begin{figure}[t]
\centering
\includegraphics[height=5.941cm,width=7.5cm]{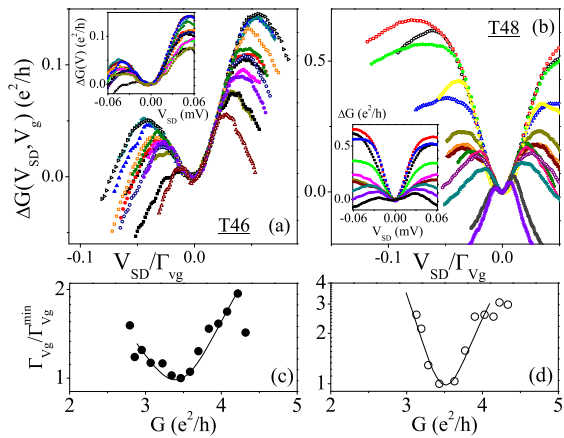}
\caption{(Color online) Scaling of the low-energy nonequilibrium
characteristics with a gate-voltage ($V_{\rm g}$)-dependent
energy-scale $\Gamma(V_{\rm g})$ at $T \approx 32$ mK for (a) T46 and
(b) T48. Insets show the unscaled, but zero-shifted, traces. Both
sets were obtained for $G$ around $4e^2/h$ in Fig.~1. (c),(d)
Variation of $\Gamma(V_{\rm g})$ as a function linear conductance $G$
in the T46 and T48 respectively. Solid lines are guide to the eye.}
\end{figure}

Alternatively, dependence of the effect on $n_{\rm s}$, {\it i.e.}
relatively strong amplitude of ZBA in samples with low-$n_{\rm s}$,
and also weakening of the effect with increasing $n_{\rm s}$ within a
given sample, indicates a possible role of Coulomb interaction.
Typically in our systems, the low magnitude of $n_{\rm s}$ ($\sim 0.8
- 3\times10^{10}$ cm$^{-2}$) results in a large interaction
parameter, $r_{\rm s} = 1/a_{\rm B}\sqrt{\pi n_{\rm s}} \sim 3.5 -
6$, where $a_{\rm B}$ is the effective Bohr radius. This corresponds
to an exchange energy that can lead to strong many-body spin
fluctuations with magnetic moment $\gg 1/2$~\cite{many_body_spin}.
Underscreening of such spin fluctuations when Kondo-coupled to lead
electrons can result in split Kondo resonance, which would be
strongly $V_{\rm g}$-dependent. Exchange splitting of Kondo-resonance
in presence of itinerant electron ferromagnetism has been observed
experimentally~\cite{pasupathy}. Here, the suppression of the double
peak pattern in $B_{||}$ provides further evidence of the dynamics to
be related to real spin, rather than pseudospin resulting from
quasi-degenerate orbital states. Indeed, similar ZBA observed in
quasi-1D quantum point contacts have been interpreted in terms of a
dynamic spin-polarization of the electrons within the 1D
channel~\cite{1D_Kondo_expt}.

Finally, we discuss the intermittent collapse of $\Delta$ observed in
Fig.~1. If the scaling relation of Eq.~\ref{scale} originates from
proximity of a 2CK fixed point~\cite{2CK_TLS_expt,ralph_PRL}, the
low-$T$ $V_{\rm g}$-dependence of $\Delta$ indicates that such
scaling should also be possible in terms of a $V_{\rm g}$-dependent
energy-scale $\Gamma(V_{\rm g})$. This implies an asymptotic form of
the scaling function ${\cal F} \rightarrow {\cal F}[V_{\rm
SD}/\Gamma(V_{\rm g})]$ as $T \rightarrow 0$. Theoretically, $\Gamma$
is analogous to the energy asymmetry in case of orbital degeneracy,
or the direct exchange parameter $I$ in the two-impurity Kondo
parametrization~\cite{Hofs_Scho,Pust_Glaz,vojta,Pust_Glaz_2004},
which depends implicitly on $V_{\rm g}$ through the exchange
interaction magnitude $J$~\cite{tarucha_J}. Indeed, in Fig.~4a and 4b
we have shown the low-energy scaling of $\Delta G(V_{\rm SD},V_{\rm
g}) = dI/dV_{\rm SD}(V_{\rm SD},V_{\rm g}) - G(V_{\rm g})$ in T46 and
T48 at the base $T \approx 32$ mK. In the absence of an absolute
scale, we have expressed $\Gamma(V_{\rm g})$ with respect to its
minimum value ($\Gamma(V_{\rm g})^{\rm min}$) observed around $G \sim
3.5\times e^2/h$ in both samples (Fig.~4c and 4d). The scalability of
$\Delta G$ close to the collapse of $\Delta$ strongly indicates the
possibility of a quantum critical dynamics. However, a satisfactory
explanation of its reentrant nature and the fundamental mechanism of
the 2CK-type scattering forms the basis of ongoing investigations.

A.G. acknowledges fruitful discussions with G. Gumbs and V. I.
Fal'ko. This work was supported by an EPSRC funded project. C.S.
acknowledges financial support from Gottlieb Daimler- and Karl
Benz-Foundation.

\end{document}